\documentclass[12pt]{article}

\usepackage{amsmath,amssymb}
\usepackage{braket}
\usepackage{graphicx}
\usepackage{color}
\usepackage{dcolumn}
\usepackage{bm}
\usepackage[numbers,super,comma,sort&compress]{natbib}

\usepackage{authblk}
\definecolor{background-color}{gray}{0.98}

\title{Quantum Monte Carlo prerequisites for industrial catalysts. Accurately assessing H atom and H$_2$ physical adsorption energy on Pt(111).}

\author{Philip E Hoggan}

\affil{Institute Pascal, UMR 6602 CNRS,BP 80026, 63177 Aubiere Cedex, France}

\begin{document}

\maketitle

\begin{abstract}
The yardstick of new first-principles approaches to key points on reaction paths at metal surfaces is chemical accuracy compared to reliable experiment. By this we mean that such values as the activation barrier are required to within 1 kcal/mol. Quantum Monte Carlo (QMC) is a promising (albeit lengthy) first-principles method for this and we are now beyond the dawn of QMC benchmarks for these systems, since hydrogen dissociation on Cu(111) has been studied with quite adequate accuracy in two improving QMC studies \cite{hog0, kdd2}.

Pt and Cu require the use of pseudo-potentials in these calculations and we show that those of Pt are less problematic than those for Cu, particularly for QMC work.

This work determines physisorption energies for hydrogen atoms and molecules on Pt(111). These systems are used as asymptotes to determine reaction barriers. As such, they must be referred to clean Pt(111) surfaces and the isolated hydrogen atom or molecule.

Previous work gave the activation barrier to hydrogen dissociation on Pt(111) using the bridging geometry. The method used is state-of-the-art ({\it ab initio}) QMC. This barrier agrees to better than chemical accuracy with a recent Specific Reaction Parameter (SRP-DFT) calculation, fitted to measurement \cite{iran}.

The results give the dissociation barrier for hydrogen on Pt(111) as 5.4 kcal/mol (QMC and 6.2 by SRP-DFT) with a QMC standard error below 0.99 kcal/mol.

This is encouraging for establishing less well-known benchmark values of industrial reaction barriers on Pt(111).
\end{abstract}


\clearpage
\makeatletter
\renewcommand\@biblabel[1]{#1.}
\makeatother
\bibliographystyle{apsrev}
\renewcommand{\baselinestretch}{1.5}
\normalsize
\clearpage

\section{\label{sec:1}Introduction}

\hskip6mm Among metal surfaces that catalyse hydrogen dissociation, Cu(111) has been the subject of very recent benchmark studies with Specific Reaction Parameter (SRP) Density Functional Theory (DFT) and state-of-the art Quantum Monte Carlo (QMC), because reliable molecular beam measurements of the late barrier for this reaction are available \cite{hog0,kdd2}.

Experiment and SRP-DFT agree with QMC. Standard error in the QMC is close to chemical accuracy (i.e. almost within a kcal/mol) and indications have been obtained in our earlier work for reducing the error \cite{kdd2}. The Pt(111) case is more favorable than copper, since the element can be described by trial wave-functions with variance an order of magnitude less than in the copper case, which is adversely affected by the presence of a full 3d shell.

Industrial catalysts often use Pt(111) for certain reactions triggered by bond-breaking. This surface also catalyses bond dissociation in hydrogen. Unfortunately, Pt(111) does not yet benefit from accurate molecular beam results for the reaction. A recent DFT study \cite{iran} shows that there are several co-ordinations of the hydrogen molecule, leading to two main categories of transitions state. The 'on-top' geometry has the molecule perpendicular to the surface plane, above a Pt atom. The reaction then appears to be almost barrier free. Such a barrier energy is inaccessible by QMC to date, being little more than the standard error in height. This work therefore focusses on the 'bridge' orientation. The agreement is very good between the DFT study and the present QMC work, with the former needing a fitted parameter to ensure that the excesses of two functionals compensate each other and the latter, described here, being completely {\it ab initio} but rather time-consuming.
The QMC methods used in this work were developed in \cite{hog0,hog1, kdd1, kdd2} where a complete description can be found.

QMC benchmarks the accurately measured and significantly higher barrier for hydrogen dissociation on Cu(111) to within some 1.5 kcal/mol \cite{kdd2}. The barrier on Cu(111) has been given by accurate molecular beam experiments at 0.63 eV or 14.5 kcal/mol.
The H$_2$ molecule dissociation on Pt(111) has a low experimental barrier which depends strongly on the molecular orientation and active site and ranges from some 0.06eV to 0.420eV. The higher barriers are for trigonal sites on the surface with minimal rearrangements and no defects \cite{iran}.

For a H$_2$ molecule parallel to the surface, the lowest experimental barrier is 0.27 eV.

Evaluating the bridging hydrogen dissociation barrier on Pt(111) is therefore likely to pose a stiff challenge to QMC. The lowest QMC standard error is just over 0.043eV (which is termed chemical accuracy, i.e. error bars within 1 kcal/mol).  This work shows hydrogen dissociation QMC barriers are reliable on Pt(111) for 'bridge' orientations (standard error under 0.043 eV or 0.99 kcal/mol). For this barrier, the DMC value is obtained here as 5.4 kcal/mol or 0.23 eV, within standard error of experiment (0.27 c.f. 0.23 $\pm$ 0.043 eV).

The present work tests interactions between physically adsorped species and Pt(111). This interaction is generally so small as to fall within standard error of the clean Pt(111) slab calculation in QMC. Nevertheless, it is useful to compare the clean surface and isolated reactant reference to the reference with all atoms involved in a reaction adsorbed without binding strongly to Pt. The purpose of this work is to test model species like hydrogen atoms and molecules to check whether the clean Pt(111) and isolated molecule would be as reliable a reference for hydrogen dissociation on platinum as the equilibrium molecule 8 \AA \hskip2mm above the surface appears to be from our previous work on H$_2$ dissociation at Pt(111) \cite{hog11}.

Any differences found, regarding the suitability of a clean surface and isolated molecule, as opposed to the physisorped molecule asymptote (equilibrium geometry at long range from the surface, will be analysed so that best practices may be justified for these heterogeneous systems in QMC. Nominally, little or no difference is expected between energies of the separate and physisoped systems. Wave-function nodes will, however be modified and the influence of this through the fixed-node approximation in QMC will be discussed.

\section{\label{sec:2} Soft pseudo-potentials and correlation for Pt atoms.}

\hskip5mm The atomic cores are represented by pseudo-potentials. Therefore, wave-function nodes cannot be exact, since these pseudo-potentials are constructed with radially node-less cores. This approximation is driven by the applications requiring accessible computer time. In spite of input node error and the absence of nuclear cusp, these pseudo-potentials are relativistically correct, which is a definite pre-requisite. The Pt atom here has a relativistic Z=60 core defined by pseudo-potentials designed by the Stuttgart group \cite{hog1} (and references therein).
\vskip2mm
\newpage

A Jastrow factor is pre-multiplied into the slater-determinant(s) constituted by Kohn-Sham spin-orbitals in the present work. This must be physically correct, in accounting for all possible interactions of electrons with different spin. The vast majority of correlation energy is accounted for by the region immediately surrounding atoms, even when few electrons are treated explicitly i.e. for large core pseudo-potentials. The spin behaviour must be carried over correctly to polyatomic systems, that is, when we consider radical species (like those containing H. radicals, below) each determinant must be pre-multiplied by a Jastrow factor with physical spin dependence. The Jastrow optimisation must be done carefully in all cases. It is general best practice to variationally optimise first with respect to variance (varmin). Then, more fine tuning can be applied with energy minimisation using Variation Monte Carlo (VMC) and optimisation of free parameters in the Jastrow factors expressed as polynomials in the instantaneous inter-particle separations.
\vskip2mm
The atomic spectra are correctly reproduced by spin-polarised guiding wave-functions in the Diffusion Monte carlo tests carried out as a prelude to this work. Note, in passing, that the platinum group metals Ni, Pd and Pt each have different atomic electronic ground state configuration. This suggests the relevance of both spin-orbit coupling NLCC pseudo-potential terms as well as the accounting of a configuration interaction of at least 3 states: Ni 3d$^8$ 4s$^2$ ground state (comparable to copper Cu 3d$^{10}$ 4s$^1$), Pd 4d$^{10}$ and Pt 5d$^9$ 6s$^1$.

Previous studies have shown that the non-locality approximation is poorly satisfied for Cu (especially with the $l=0$ local channel but also with other choices. Tests have shown that this is a minor problem for Pt, particularly with the $l=3$ choice.

Resolution of the electronic spectrum for the Pt atom also shows accurate excitation energies for spin-polarised cases, in particular for the transition between the ground-state Pt 5d$^9$ 6s$^1$ (GS $^3D_3$) and Pt 4d$^{10}$. $^1S_0$

The pseudo-potential supplied with ABINIT (FHI 10 active electrons) gives :

GS at -26.0113195 Ha

ES at -25.9833442 Ha (large-CI value -25.9851687 Ha, from \cite{CI}; a 6\% error in barrier-energy.

Transition: 0.762 eV (error 0.0496eV) Using the Stuttgart PP (60 electron core $\it vide infra$) gives 0.76 eV (exact to 0.25 \%).

The other excited states are resolved quite adequately.

The fixed-node error arises because QMC uses a product of input wave-function and its update to represent 'density' and even the accurate Diffusion step (DMC) where configurations are updated to represent this density will have the same nodes. A significant portion of the fixed-node error is due to atomic cores. It is minimised by subtracting energies corresponding to structures with the same number of metal atoms. This feature is also observed when comparing clean metal surfaces and gaseous reactant with an atom or gas molecule physisorped at the metal surface, as a precursor for the reacting system. This resolution has been tested successfully in the cases of hydrogen radicals as well as the H$_2$ molecule physisorped on Pt(111).

\section{\label{sec:3} Results for hydrogen atoms and molecules on Pt(111).}

The H-atom adsorption was tested without recourse to Casula t-moves.

We obtain: (with l=3 as pseudo-potential local channel).

Clean Pt(111) surface: -409.1354 $\pm$ 0.002

Surface with a H atom: -409.6360 $\pm$ 0.003

Difference: -0.5006 $\pm$ .0036. (the exact isolated H. is -0.5 by definition and some very small stabilisation estimated at .0004 Ha may be deducted).

Solid models expose Pt(111) and are described in detail in \cite{hog11}. They are generally 5 layers thick and most studies use a 2 by 2 mesh (with some double sized cases). The clean surface averages -408.428 Ha and that with physisorped H$_2$ -409.687. This implies that resolution of the H$_2$ energy requires twist-averaging and various corrective terms, as the free molecule can be obtained with exact total energy at -1.1758 Ha. Note that the optimised Variation Monte Carlo (VMC) values, using energy minimisation give brute differences close to the exact value: clean Pt(111) at -407.56 and -408.73 with physisorped H$_2$. The difference is 1.17 Ha. Since the standard error for this test is close to 0.01 Ha, it is essentially exact. This hints on a defect in the pseudo-potential that becomes apparent when the more accurate Diffusion Monte Carlo (DMC) method is applied. The Pseudo-potential fails to satisfy the locality approximation. This has been well-documented for copper in our previous work \cite{kdd1}. It can be shown to cause difficulties for physisorption of hydrogen by using Casula t-moves and by varying the local channel of the pseudo-potential (one of the harmonics of order l is chosen to be local, generally the highest present, l=3 here, but tests with the highest occupied (l=2) and the lowest (l=0) have been summarised below. This accounts for part of the defect in resolving the H$_2$ physisorption. The remainder can be resolved by comparing the same geometries with the proviso that the hydrogen be elongated (in the transition-state geometry) or at the equilibrium value (physisorbed or even isolated hydrogen molecules).

The t-moves bring the estimate of hydrogen molecule total energy from a twist-averaged (over 20 offset grids) -1.1510 Ha to -1.1721 Ha (difference 0.0037). This is the contribution to systematic error of non-locality in the pseudo-potential. It leads to 2.5 \% error in the total energy of H$_2$, therefore not as drastic as some 3d dimer binding energy errors. Subtracting the same atoms gives the a binding energy that corresponds to -1.1760 Ha (0.0002). This is indicative of the nodes being systematically in error in the trial wave-function. The subtraction improves the fixed-node error by about 95\% in this case.

{\bf Table I Total Energy without t-moves in atomic units.}
\vskip1mm

 \begin{tabular}{|c|c|c|c|c|}
  \hline
 Structure & VMC E$_{tot}$  & DMC l=3  & DMC l=2 & DMC l=0 \\
 \hline
 Clean Pt(111) 2x2 mesh & -407.56 & -408.536 & -408.475 & -408.602 \\
\hline
Hydrogen molecule (phys.)  & -408.73 & -409.687 & -409.664 & -409.706\\
\hline
Hydrogen molecule E (tot)  & -1.17 & -1.15 & -1.189 & -1.104  \\
\hline
\end{tabular}

\vskip2mm

Note that the l=0 local channel is the worst-case scenario. The work continued with Casula t-moves and l=3 as local channel which gives the accurate atomic spectra for Pt.

\vskip2mm

{\bf Table II Total Energy (H$_2$ as difference of Pt(111) with and without it). \\ Uses t-moves in atomic units.}
\vskip2mm

 \begin{tabular}{|c|c|c|c|c|}
  \hline
 Structure & VMC E$_{tot}$  & DMC l=3  & DMC atom cons. & DMC gas \\
 \hline
Hydrogen molecule E (tot)  & -1.17 & -1.1721 & -1.1760 & -1.1758485 \\
\hline
Standard error & 0.0005 & 0.00003 & 0.00002 & 0 \\
\hline
\end{tabular}

\vskip4mm

\section{\label{sec:4} Perspectives and conclusions.}

This study shows that calculations involving the clean Pt(111) surface are consistent with those using asymptotic physisorped geometries. The 'same atom' systems are, nevertheless, more accurate. This phenomenon is ascribed to a cancellation of the majority of fixed-node error, when the nodes should be comparable. This is not the case when a clean surface is involved, because of intrinsically 2-D surface states, having specific nodal structure. This 2-D symmetry is broken once an atom is adsorbed. The comparison of systems with the same atoms further renders the nodes comparable, although not identical because of the consequences of geometry changes. The formation and dissociation of bonds is certain to influence nodal structure and this effect on fixed node error will be the subject of future work. Some authors have already shown for LiH dissociation in QMC the advantages of a valence bond CSF with breathing orbitals. We will investigate related strategies.

This work completes preparation for elementary reaction step studies in contact with Pt(111) and subsequently, the rate-limiting step of catalytic processes can be identified.

\end{document}